# Stereochemically Active Lone-pair Leads to Strong Birefringence in the Vacancy Ordered $Cs_3Sb_2Cl_9$ Perovskite Single Crystals

*Shramana Guha,[a] Amit Dalui,[a] Piyush Kanti Sarkar,[a] Sima Roy,[a] Atanu Paul,[b] Sujit Kamilya,[c] Abhishake Mondal,[c] Indra Dasgupta,[b] D. D. Sarma[c] and Somobrata Acharya[a,d]**

[a]School of Applied and Interdisciplinary Sciences, Indian Association for the Cultivation of Science, Jadavpur, Kolkata-700032, India.

[b]School of Physical Sciences, Indian Association for the Cultivation of Science, Jadavpur, Kolkata-700032, India.

[c]Solid State and Structural Chemistry Unit, Indian Institute of Science, Sir C V Raman Road, Bangalore 560012, India.

[d]Technical Research Center, Indian Association for the Cultivation of Science, Jadavpur, Kolkata-700032, India.

*E-mail: camsa2@iacs.res.in

ABSTRACT. Stereochemically active lone-pair (SCALP) cations are attractive units for realizing optical anisotropy. Antimony (III) chloride perovskites with SCALP have remained largely unknown till date. We synthesized vacancy ordered $Cs_3Sb_2Cl_9$ perovskite single crystals with $SbCl_6$ octahedral linkage containing SCALP. Remarkably, $Cs_3Sb_2Cl_9$ single crystals exhibit an exceptional birefringence of 0.12 ± 0.01 at 550 nm, which is the largest among pristine all-inorganic halide perovskites. The SCALP brings a large local structural distortion of the $SbCl_6$ octahedra promoting birefringence optical responses in $Cs_3Sb_2Cl_9$ single crystals. Theoretical calculations reveal that the considerable hybridization of Sb 5*s* with Sb 5*p* and Cl 3*p* states largely contribute to the SCALP. Furthermore, the change in the Sb-Cl-Sb bond angle creates distortion in the $SbCl_6$ octahedral arrangement in the apical and equatorial directions within the crystal structure incorporating the required anisotropy for the birefringence. This work explores pristine inorganic halide perovskite single crystals as a potential birefringent material with prospects in integrated optical devices.

**TOC GRAPHICS**

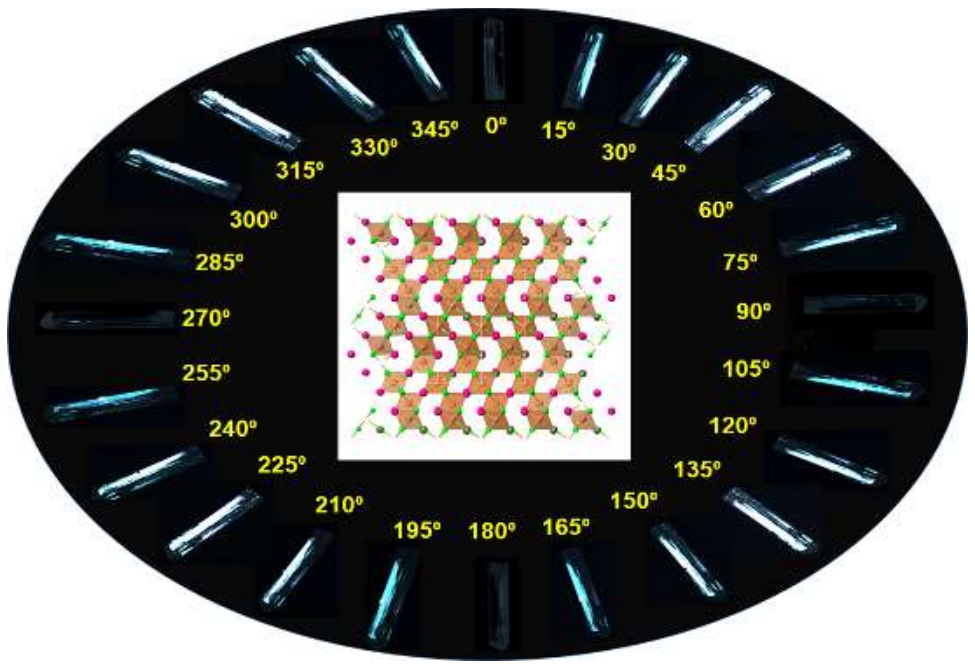

Birefringent crystals contain difference in the refractive indices along the different crystal axes. Optical anisotropy of a birefringent crystal splits the incident light into two beams with orthogonal oscillations termed as the ordinary ray (o-ray) and the extraordinary ray (e-ray).[1] Birefringent compounds play an indispensable role in modern optoelectronics and information technology.[2] Optical components such as optical circulator,[3–5] beam splitters,[4,6–9] electro-optics Q switch,[4] optical isolators,[5,10] achromatic quarter waveplates[1,10,11] and phase compensators[1,10,11] are realized using birefringent crystals. Currently, non-centrosymmetric oxides have been extensively investigated as outstanding birefringent materials.[7,12,13] Oxide birefringent crystals such as $TiO_2$,[14] $YVO_4$,[15] $LiNbO_3$,[16] $CaCO_3$,[17] and $\alpha$-$BaB_2O_4$,[18] with birefringence ranging from 0.256 to 0.122 in the visible range are used commercially.[10,19] In spite of few exceptions[11,20–25] birefringent materials are almost exclusively limited to oxides till date. Recently, halide perovskites emerged as potential energy materials owing to the impressive optical and electronic properties.[26–32] In comparison, the possibility of halide perovskites as birefringent materials has long been overlooked. Recently, birefringence was observed from the mixed halide or hybrid halide perovskite materials.[1,10,33] Chen *et al*. reported the fabrication of achromatic quarter-wave plates using inorganic $Cs_4PbBr_6$ crystals embedded with perovskite $CsPbBr_3$ nanocrystals.[1] The planar π-conjugated melamine contained layered hybrid halide perovskite $MLAPbBr_4$ showed birefringence of 0.322 at 550 nm, which is higher than the commercially available oxide birefringent crystals.[10] These results demonstrated that the contribution of A-site cations to the birefringence was less due to their almost sphere-like electron distribution while anisotropic response of the orbitals on the B-site cation and X-site anion groups to incident light contributed to the birefringence.[19,34] However, these pioneering reports incorporated inorganic mixed crystals or organic melamine cation along with halide perovskites, while the birefringence of any all-inorganic pristine halide perovskite is not reported yet.

All inorganic $A_3B_2X_9$ (A= $NH_4$, Tl, K, Rb, Cs; B = Sb, Bi; X = Cl, Br, I) perovskites appear as emergent alternative of $ABX_3$ perovskites due to the analogous electronic configuration.[35] Various vacancy ordered $A_3B_2X_9$ structures can be regarded as defect variants of cubic $ABX_3$ perovskite-type involving three formula units of generating parent cubic $ABX_3$ structure.[36,37] However, compared to the $ABX_3$ structure, the vacancy ordered $A_3B_2X_9$ perovskites permit more alternatives for octahedral linkage.[36] $Cs_3Sb_2X_9$ perovskites exist in the trigonal or orthorhombic crystal phases where trigonal phase contains less crystal packing density compared to the orthorhombic phase.[35] These halide perovskites consist of $SbX_6$ polyhedra, which are corner shared to form a layered structure in the trigonal phase, while these perovskite crystals form zigzag double chains in the orthorhombic phase.[38] Metal cations with stereochemically active lone-pair (SCALP) greatly contributed to the birefringence by causing distortion in the structural units incorporating the optical anisotropy for the vast majority of birefringent compounds.[19,21,25,39–44] The mechanism of SCALP in antimony halides has been a popular issue since SCALP of the $SbX_6$ containing perovskites remains controversial. Considerations based on the crystal structures encounters certain ambiguity in determining the character of the SCALP. For example, earlier studies demonstrated $SbX_6$ polyhedra with perfectly octahedral structures to be stereochemically inactive lone-pair.[45] Recent electronic structure calculations revealed stereochemical activity of antimony cations in antimony halides, phosphates and chalcogenide compounds.[2,21,39] Hence, the occurrence of the SCALP on the metal cation of $Cs_3Sb_2X_9$ perovskites should considerably influence the geometry of the $SbX_6$ polyhedra and resultant birefringence. Besides the achievement of SCALP in Sb cations, another important issue is to arrange the multiple SCALP cations in an acentric structure to adopt global asymmetric environments. Here we report on the synthesis of $Cs_3Sb_2Cl_9$ perovskite single crystals with SCALP containing $SbCl_6$ octahedral framework *via* a solvothermal method. Millimetre sized ($8 \times 2 \times 1$ mm$^3$) $Cs_3Sb_2Cl_9$ single crystals in the orthorhombic crystallographic phase show large birefringence of 0.12 ± 0.01 at 550 nm. This outstanding birefringence performance of the halide perovskite single crystal is comparable to the commercially available oxide birefringent crystals. We show that the hybridization of Sb 5*s* with Sb 5*p* and Cl 3*s* states determines the SCALP of antimony cations. Our findings indicate that SCALP changes the classical Sb coordination octahedra into distorted polyhedra. The change in the Sb-Cl-Sb bond angle also imparts distortion in the $SbCl_6$ octahedral arrangement in the apical and equatorial directions within the crystal structure imparting the required anisotropy for the birefringence. To the best of our knowledge, herein, we are reporting the first example birefringence of a pristine all-inorganic halide perovskite single crystals.

Slow cooling technique is a solution phase method where saturated precursor solution at high temperature is slowly cooled down to room temperature to grow the single crystals.[46] We synthesized $Cs_3Sb_2Cl_9$ single crystals by adopting the slow cooling technique using the solution of $SbCl_3$ (0.457 g, 2.0 mmol) in 37% HCl with a 1 mL solution of CsCl (0.504 g, 3.0 mmol) in 37% HCl under stirring. After 60 minutes of vigorous stirring at room temperature, a white precipitate was obtained by re-crystallization in HCl. A dilute solution (0.05 g/mL) of the white precipitate in 37% HCl was prepared and heated at 120°C for 60 minutes to make a clear solution. Then, slow cooling (1°C/ 10 min) of this solution over 10 hours led to the formation of whitish rod-shaped $Cs_3Sb_2Cl_9$ single crystals with dimensions of $8 \times 2 \times 1$ mm$^3$ (Figure 1a). Single crystal X-ray diffraction (SCXRD) measurements were performed to extract the crystal structure of the as-synthesized $Cs_3Sb_2Cl_9$ crystals (Figure 1b). Details of the crystallographic parameters obtained from crystallographic information file (CIF file) are tabulated in table S1. SCXRD reveals orthorhombic centrosymmetric crystal phase ($\beta$-phase, $a$ = 18.65 Å, $b$ = 7.57 Å, $c$ = 13.06 Å, $\alpha = \beta = \gamma = 90°$) with space group of P*nma*. We have measured lattice plane orientation of a $Cs_3Sb_2Cl_9$ single crystal by measuring SCXRD on top of the planar surface of a single crystal, which reveals (114) lattice planes perpendicular to the elongated surface of the rod (Figure S1). The SCXRD pattern also reveals the highest intensity of the (114) planes suggesting that preferred growth direction is along the elongated surface of the rod (Figure 1b). Furthermore, the SCXRD data was integrated to obtain the powder XRD pattern, which matches quite impressively with the experimentally obtained powder XRD pattern (Figure 1b), suggesting that $Cs_3Sb_2Cl_9$ retains the orthorhombic phase in its microstructures.

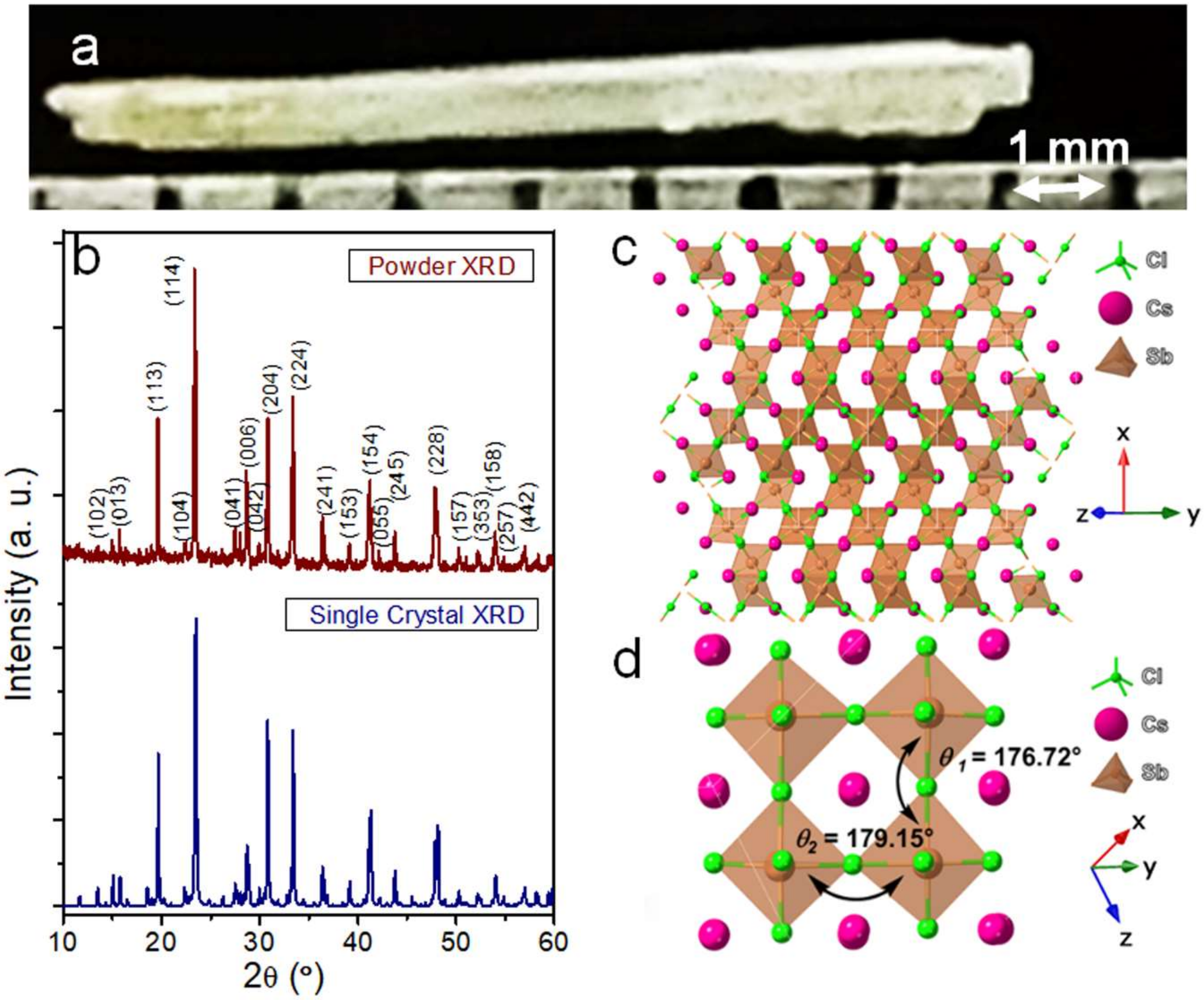

**Figure 1.** (a) Photograph of a $Cs_3Sb_2Cl_9$ perovskite single crystal. (b) Powder XRD and single crystal XRD of $Cs_3Sb_2Cl_9$ single crystals showing orthorhombic crystal phase. (c) Crystal structure of the vacancy ordered $Cs_3Sb_2Cl_9$ single crystals showing zigzag double chains in the orthorhombic phase. (d) The average Cl-Sb-Cl bond angles along the apical and equatorial planes showing the larger equatorial bond angle than the apical direction.

Further dilution (0.04 g/ mL) of the pristine white precipitate in HCl and heating at 120°C for 60 minutes followed by slow cooling (1 °C/10 min) over the course of 10 hours resulted in large single crystals of hexagonal shape (Figure S2). These hexagonal shaped single crystals were grown in the trigonal crystal phase instead of orthorhombic phase (Figure S2). The use of little higher concentrated solution (0.07 g/mL) led to the formation of colorless crystalline chunks under the same reaction conditions (Figure S2). These crystalline chunks showed a mixture of orthorhombic and trigonal crystal phases (Figure S2). On the other hands, fast cooling of the solution (0.05 g/mL) under the same reaction condition resulted in the formation of irregular plate shaped single crystals with orthorhombic crystallographic phase (Figure S2). Thus, it is evident that the morphology and even the crystallographic phase of the resulting single crystals are sensitive to the synthesis conditions.

The structure of rod-shaped $Cs_3Sb_2Cl_9$ single crystals consists of Cs atoms and interconnected $SbCl_6$ octahedra (Figure 1c). The asymmetric units contain three Cs atoms, two Sb atoms and six Cl atoms (Figure S3). In the asymmetric unit, Sb1 is bonded with four Cl atoms (Cl1, Cl2, Cl3 and Cl4), and Sb2 is bonded with three Cl atoms (Cl4, Cl5, and Cl6) where both $SbCl_6$ polyhedral are completed by symmetrically related Cl atoms (Cl4, Cl1 for Sb1 and Cl1, Cl4, Cl6 for Sb2). The Sb atoms (Sb1 and Sb2) are interconnected by two Cl atoms (Cl4 and Cl1) to form an Sb-Cl-Sb square, and each square is interconnected in a zigzag manner forming 1D chains (Figure 1c and Figure S3). Two Cs atoms are positioned between two 1D chains interacting with six nearby Cl atoms. The terminal Sb1-Cl bond distances are 2.505 Å (Sb1-Cl2) and 2.502 Å (Sb1-Cl3), whereas the bridging Sb-Cl bond distances are 2.811 Å (Sb1-Cl1) and 2.799 Å (Sb1-Cl4) respectively. For Sb2-Cl bonds, the terminal Sb-Cl bond distances are 2.462 Å (Sb2-Cl6) and 2.448 Å (Sb2-Cl5) and the bridging Sb-Cl bond distance is 2.856 Å (Sb2-Cl4). Thus, the bridging Sb-Cl bond distances are longer than the terminal Sb-Cl bond distances for both the Sb atoms, which is a signature of existence of Sb (III) oxidation state. The average Cl-Sb-Cl bond angles are different along the directions of either apical (173.56°) or equatorial plane (177.74°) respectively (Figure 1d). The bond angle of equatorial Cl-Sb-Cl in $Cs_3Sb_2Cl_9$ orthotropic phase is larger by at least 4° than that of apical one.

The single crystals were grinded finely for the transmission electron microscope (TEM) measurements (Figure S4a). The high-resolution TEM (HRTEM) image (Figure 2a) reveals an interplanar spacing of ~0.38 ± 0.05 nm corresponding to the (114) planes of bulk orthorhombic crystal structure (ICSD 01-074-4795). Selected area electron diffraction (SAED) pattern also shows diffraction spots corresponding to the orthorhombic phase (Figure 2b). Energy dispersive X−ray spectroscopy (EDS) measurements in TEM reveals the atomic ratio of Cs:Sb:Cl ≈ 3:2:7, which closely matches with $Cs_3Sb_2Cl_9$ chemical composition of the crystals (Figure S4b). HAADF-STEM image and EDS elemental mapping shows homogeneous distribution of the constituting elements (Figure S4c-e). The thermogravimetric analysis curve reveals that $Cs_3Sb_2Cl_9$ single crystals decompose above 300 °C retaining 51% of the initial weight up to 700 °C (Figure S5a). No phase transition was observed in the differential scanning calorimetry measurements before the decomposition temperature (Figure S5b). The single crystals are stable over two months at 55% relative humidity or under UV irradiation for 1 hour supporting ambient stability in comparison to most of the halide perovskite crystals.

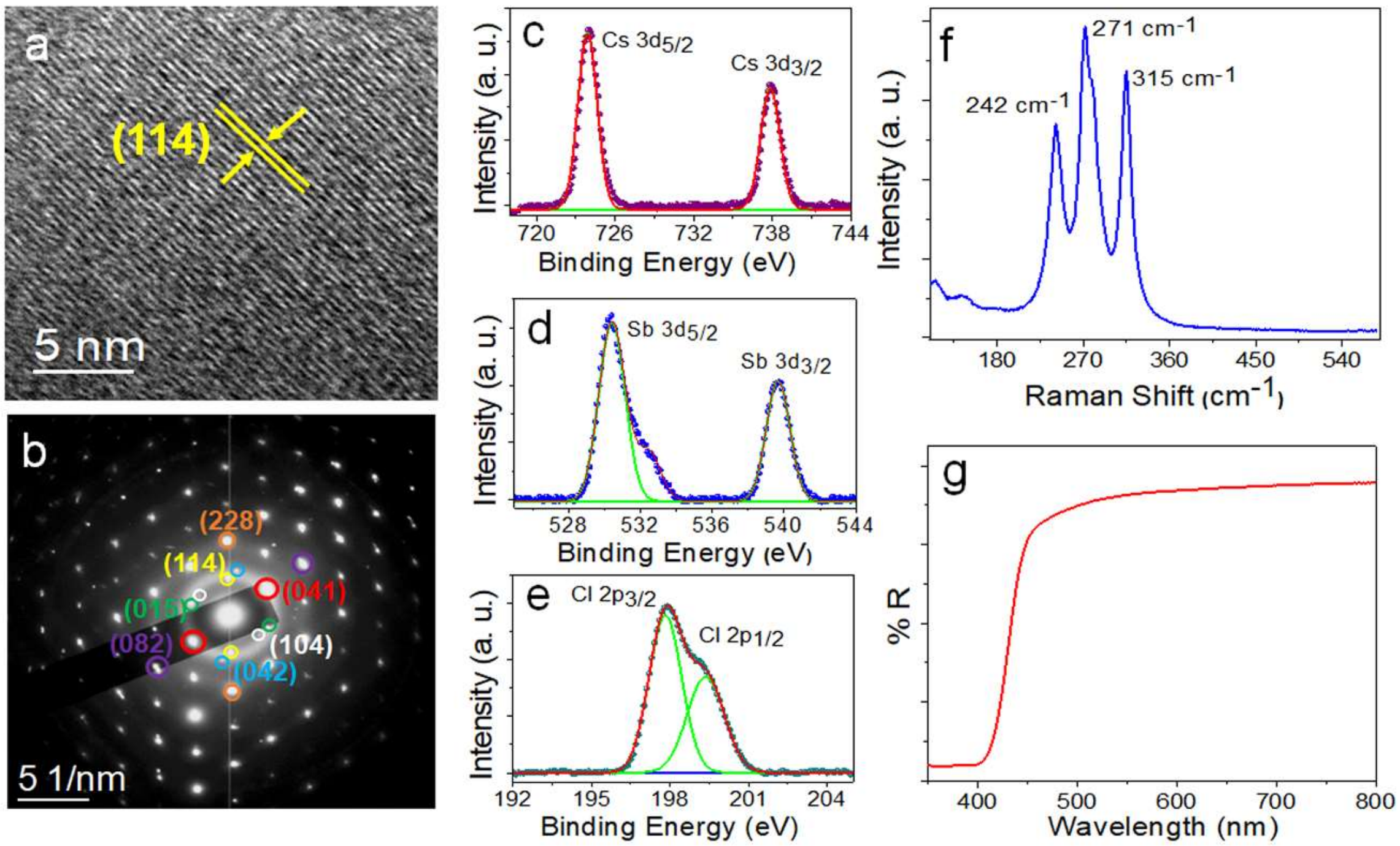

**Figure 2.** (a) HRTEM image of grinded $Cs_3Sb_2Cl_9$ single crystals showing interplanar lattice spacings corresponding to (114) planes of orthorhombic crystal structure. (b) SAED pattern of grinded $Cs_3Sb_2Cl_9$ single crystals showing diffraction spots corresponding to orthorhombic crystal structure. High resolution XPS spectra of (c) Cs 3*d* orbitals (d) Sb 3*d* orbitals and (e) Cl 2*p* orbitals. (f) Raman spectrum of $Cs_3Sb_2Cl_9$ single crystals showing distinct peaks at 242 $cm^{-1}$, 271 $cm^{-1}$ and 315 $cm^{-1}$ respectively. (g) UV-vis diffuse reflectance spectrum of $Cs_3Sb_2Cl_9$ single crystals.

We measured X-ray photoelectron spectroscopy (XPS) to probe the oxidation states of the elements of $Cs_3Sb_2Cl_9$ single crystals (Figure S6). The energy positions of the respective features clearly indicate the expected charge states of all the elements. XPS spectrum confirms the presence of Cs (+1), Sb (+3) and Cl (-1) oxidation states of the elements respectively in the single crystals.

XPS spectrum reveals well-resolved peaks at binding energies of 737.9 eV and 723.9 eV corresponding to Cs $3d_{3/2}$ and Cs $3d_{5/2}$ respectively (Figure 2c). Sb element has two peaks located at 539 and 530 eV, which are assigned to Sb $3d_{3/2}$ and Sb $3d_{5/2}$, respectively (Figure 2d). The peaks at 199.21 eV and 197.82 eV can be assigned to the Cl $2p_{1/2}$ and Cl $2p_{3/2}$ binding energy respectively (Figure 2e).[35] Detail analyses of XPS spectra reveal a relative elemental ratio of Cs:Sb:Cl ≈ 3:2:8 which matches closely with the EDS measurements in TEM (Figure S4). In the orthorhombic phase, $SbCl_6$ octahedra with a lower symmetry ($C_{2v}$) contains two different types of Sb-Cl bonds and each octahedra contains four types of Sb-Cl bonds, as evidenced from the single crystal structure analyses and further confirmed by Raman spectroscopy. Raman spectrum of the $Cs_3Sb_2Cl_9$ single crystals at room temperature shows three typical bands at 242, 271, and 315 $cm^{-1}$ respectively corresponding to the Sb-Cl bond vibrations (Figure 2f). Raman bands can be assigned to stretching of antimony with the bridged and terminal chlorines moieties.[38] The UV-vis diffuse reflectance spectrum of $Cs_3Sb_2Cl_9$ crystals shows no obvious absorption peak in the visible region (Figure 2g). However, a strong absorption in the UV region is evidenced suggesting visible transparent nature of the single crystals. Band gap obtained using Tauc plot indicates a direct band gap of 2.98 eV for $Cs_3Sb_2Cl_9$ single crystals (Figure S7).

Polarization resolved optical microscopy (PROM) is an effective tool which is often used to determine the birefringence properties.[44,47] We investigated the birefringence properties of $Cs_3Sb_2Cl_9$ single crystals using PROM. In the experimental set-up, the polarizer and the analyzer of the PROM were set perpendicular to each other, and a single crystal was mounted on the circular microscope rotating stage. The crystal was rotated through 360° with a step of 15° while maintaining the crossed polarization conditions (Figure 3a). The single crystal appeared periodically bright and dark with the rotation of 45°. It was observed that the single crystal appeared the brightest for 45°, 135°, 225° and 315° and the darkest for 0°, 90°, 180°, 270° and 360° (Figure 3a). This observation proves the birefringent nature of the $Cs_3Sb_2Cl_9$ single crystals. The origin for this phenomenon is the refraction of the incident linearly polarized light into two components of lights with orthogonal vibration directions while passing through a birefringent crystal.[12,48] A resultant part of these two components of the refracted light pass through the analyzer to display birefringence upon rotation of the crystal giving the brightness. The value of birefringence was measured using a pre-selected thin crystal plate of $Cs_3Sb_2Cl_9$ single crystal under the polarizing microscope by inserting a Berek compensator.[10] The original interference color of the $Cs_3Sb_2Cl_9$ crystal under orthogonal polarized light is second-order cyan (Figure 3b). A complete extinction was achieved by inducing retardation using the compensator (Figure 3c). Simultaneously, the background changes from black to cyan showing the interference pattern.[10,20] The average thickness of $Cs_3Sb_2Cl_9$ specimen measured using scanning electron microscopy (SEM) is 6.91 μm (Figure S8) and the optical path difference measured using the compensator is 0.83 μm at 550 nm. Using these experimental parameters, we have obtained the birefringence of $Cs_3Sb_2Cl_9$ crystal to be 0.12 ± 0.01 at 550 nm (for birefringence calculations see the Supporting Information).[10,49,50] To the best of our knowledge, this is the best report on the birefringence from inorganic halide perovskite single crystals. The measured birefringence is comparable to the hybrid halide $MLAPbBr_4$ perovskite and commercial oxide birefringent crystals.[9,10,19] It is noteworthy to mention that the perovskite materials are susceptible to distortions, which contributes to the optical properties and birefringence. In general, birefringence is closely related to the orientation of the octahedral distortion and relative orientation of A-site ionic groups when hybrid halide perovskites are considered. The delocalization of A-site organic groups along with the distorted octahedral arrangement contributes to the polarizabilities facilitate the birefringence of the hybrid halide perovskites. The birefringence of $MLAPbBr_4$ originated due to the delocalization of π-conjugated melamine cations and distorted $PbBr_6$ octahedra, while $Pb^{2+}$ remained stereochemically inert.[10] In contrary, the birefringence of all-inorganic $Cs_3Sb_2Cl_9$ perovskite single crystals originated due to the SCALP of the $Sb^{3+}$ lone-pair which creates the distortion of the $SbCl_6$ octahedral units while the contribution of the $Cs^+$ to the birefringence is little comparison to the organic MLA. Additionally, the hybrid halide perovskites are prone to degrade and the organic components are quite lossy in comparison to all-inorganic halide perovskites. In the all-inorganic single crystals with significantly fewer defects and lesser inner grain boundaries, these obvious challenges are avoidable making $Cs_3Sb_2Cl_9$ perovskite single crystals a significant birefringent material.

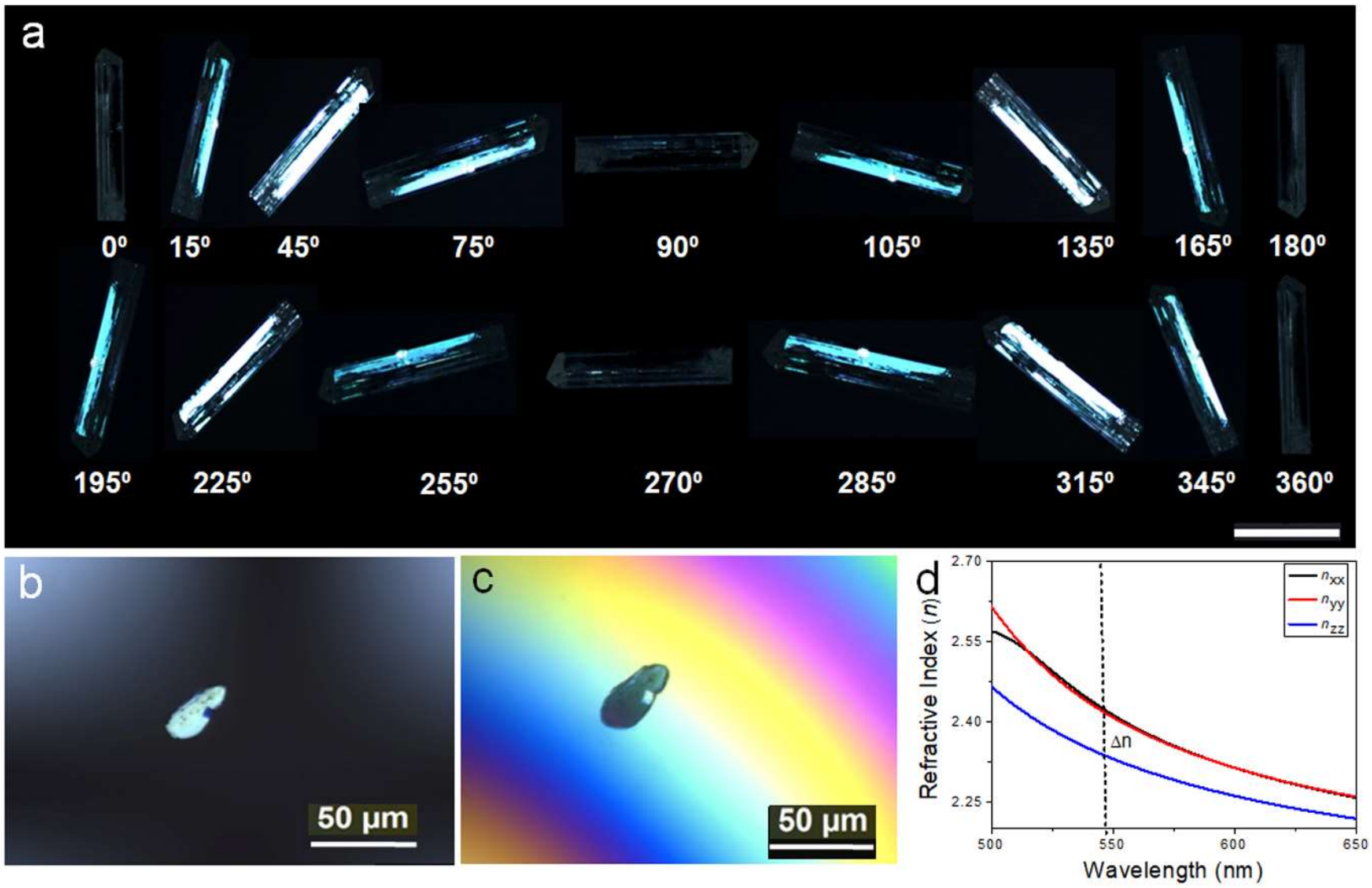

**Figure 3.** (a) Polarization resolved optical images under crossed-polarized light illumination in the PROM. The polarization direction of the incident light and the polarization analyzer are perpendicular to each other with the $Cs_3Sb_2Cl_9$ single crystal rotation angle from 0° to 360°. The step of rotation angle is 15°. (b) Original interference color of the $Cs_3Sb_2Cl_9$ crystal under crossed-polarized light. (c) The $Cs_3Sb_2Cl_9$ crystal under complete extinction. (d) The calculated wavelength dependent refractive indices of $Cs_3Sb_2Cl_9$ crystal.

Since the $Cs_3Sb_2Cl_9$ single crystals exhibit an anisotropic linear optical response, there exists unequal principal refractive indices along the principal optical axes. The difference of refractive indices reflects the birefringence for the crystal system. The inter-band transitions important for optical activity in the low energy range primarily involves the Sb-*s*, Sb-*p* and Cl-*p* states. In a crystal structure with orthorhombic symmetry, three different refractive indices are associated due to the three mutually orthogonal axes. Calculated refractive indices using density functional theory (DFT)[51,52] show $n_x$ is almost equal to the $n_y$ while $n_z$ is different (Figure 3d). Wavelength dependent refractive indices reveal a strong anisotropy with a birefringence of $\Delta n$ = 0.08 at 550 nm, which is in good agreement with the measured birefringence. To elucidate the mechanism of the large birefringence of $Cs_3Sb_2Cl_9$ single crystals, we performed first-principles electronic structure calculations in the framework of DFT[51,52] using the plane-wave basis set supplemented with projector augmented wave method (PAW) as implemented in the Vienna ab-initio simulation package (VASP) (for computational details see the Supporting Information).[53–55] The total as well as the partial DOS projected on Sb-*s*, Sb-*p*, and Cl-*p* states are shown in Figure 4a. The Cl-*p* and Sb-*s* states are fully occupied with the latter lying lower in the energy. These Cl-*p* and Sb-*s* orbitals hybridize to form a pair of occupied bonding and antibonding states. The Cl-*p* and Sb-*s* further hybridize with the empty Sb-*p* states. The electronic structure is consistent with the nominal ionic formula $(Cs^{+})_3(Sb^{3+})_2(Cl^{-})_9$ suggesting the existance of Sb-*s* lone-pair. The band gap calculated from the total DOS is 2.36 eV which is smaller in comparison to the experimental band gap as expected in a DFT calculation.

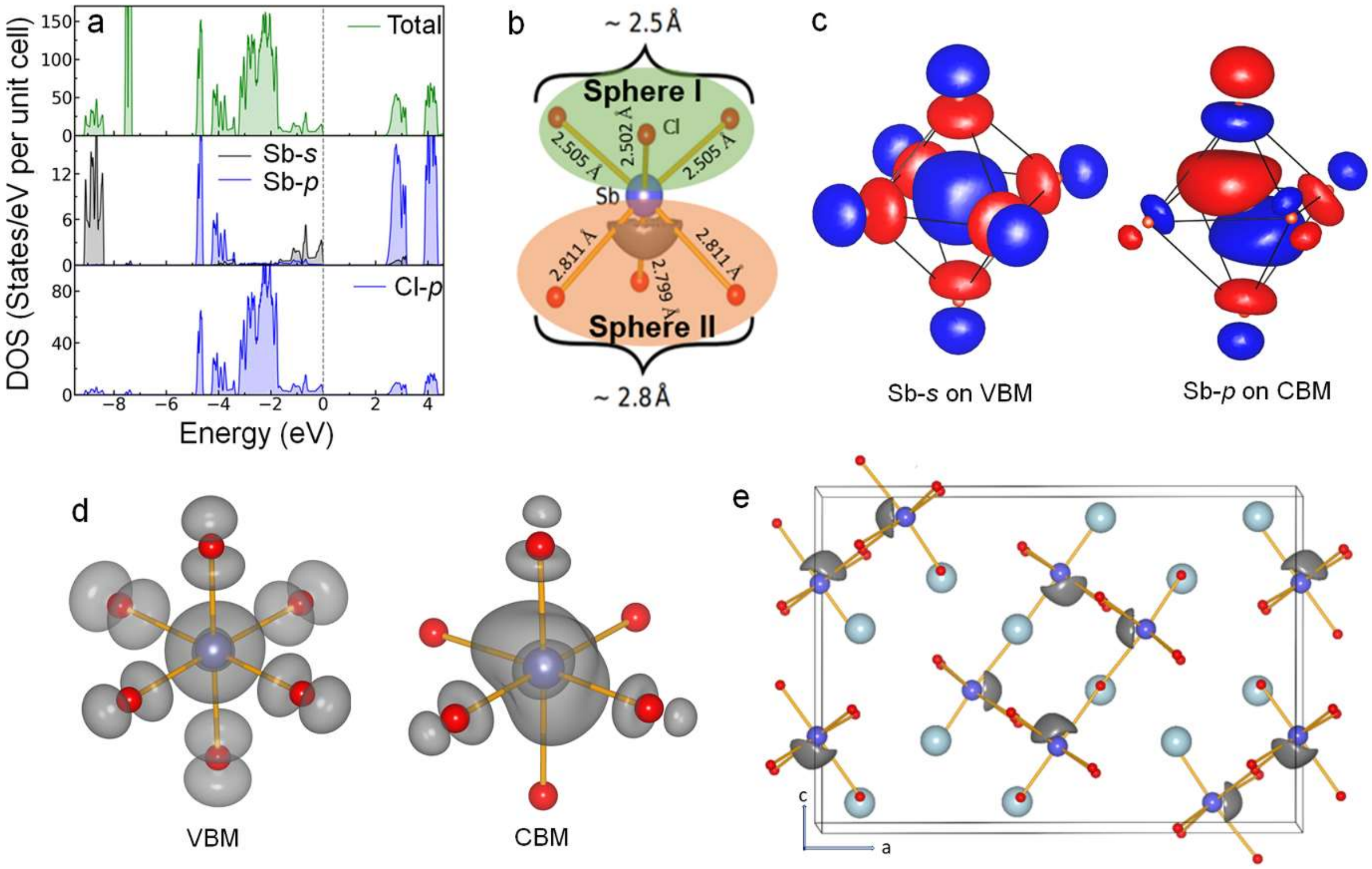

**Figure 4.** (a) Total DOS and partial DOS of $Cs_3Sb_2Cl_9$ crystals. (b) Octahedra of stereochemically active lone-pair containing cation (Sb) with anion (Chlorine). Blue ball is the cation and red balls are anions respectively. Lobe like structure (gray) is the stereochemically active lone-pair in the hemisphere- II. (c) Wannier function of Sb-*s* as projected on VBM and Sb-*p* as projected on CBM. (d) Charge density plots of VBM and CBM of $SbCl_6$ octahedra. (e) ELF diagram for $Cs_3Sb_2Cl_9$. Blue, red and cyan balls represent Sb, Cl, and Cs atoms respectively. Lone pairs on Sb (III) are shown in gray.

$Cs_3Sb_2Cl_9$ in the orthorhombic phase at room temperature is centrosymmetric. The valence state of the Sb atoms considerably influences their stereochemical role in the crystal structure. Usually Sb (V) atoms, in absence of any lone pair electrons, form stereochemically inactive equivalent Sb−Cl bonds with chlorine atoms in an octahedral coordination. However, in the present case, Sb (III) atom forms a variety of coordination polyhedra with nonequivalent Sb−Cl bonds (Figure 4b). Because of the low symmetry, both bond lengths of Sb−Cl and bond angles of Sb−Cl−Sb are different. In the distorted octahedral structures, Cl atoms are usually located on one side of the Sb (III) at relatively short distances (hemisphere I), and the other Cl atoms are located on the other side (hemisphere II) at relatively longer distances. This local distortion of the Sb (III) coordination polyhedra occurs due to the effect of the lone-pair electrons localized on the Sb (III) atom in the hemisphere II (Figure 4b). The stereochemical activity of lone-pair active Sb (III) atoms originate from the hybridization between 5*s*, 5*p* orbitals at the Sb (III) site and 3*p* orbitals at the Cl site. The 5*p*-3*p* hybridization turns out to be the governing factor in giving rise to the directionality of the lone-pair and driving the off-centric distortion of the unit cell.[56] This is further corroborated by the Wannier function plots (Figure 4c). The Wannier function for the valence band maximum (VBM) projected on the Sb-*s* states shows a strong hybridization with Cl-*p* states while the Wannier function of the conduction band minima (CBM) projected on the Sb-*p* state reveal strong hybridization with the Cl-*p* states. As a consequence, both the Sb-*s* and Sb-*p* states are involved in the formation of asymmetric charge distribution where empty Sb-*p* states are able to interact due to the presence of Sb-*s* and Cl-*p* occupied antibonding states emphasizing the importance of Cl-*p* states in the formation of lone-pairs leading to the stereochemical activity. The 5*p*-3*p* hybridization turns out to be the key factor for the generation of directionality of the lone-pairs.[57] The degree of stereochemical activity of the lone-pair may be measured by calculating the difference between the Sb−Cl bond lenghts from hemisphere I and II (Figure 4b), where $\mathbf{\Delta}d_1$ = 0.003 Å and $\mathbf{\Delta}d_2$ = 0.012 Å specify stereochemical activity of the $Sb^{3+}$ lone-pair in the crystal structure. The charge density corresponding to the VBM and CBM further corroborate asymmetric charge distribution particularly for the CBM contributing to the birefringence of $Cs_3Sb_2Cl_9$ single crystals (Figure 4d).

Exploring the contributions of microscopic units that form the perovskite framework is one efficient way to design new materials with birefringence performance. Electron localization function (ELF) is a useful method to describe localized electronic distribution like asymmetric lone-pair electronic distribution. As mentioned, the presence of Sb in a +3 oxidation state suggests the possibility of the stereochemical activity of a Sb lone-pair formed from $5s^2$ electrons.[58] The lone-pairs arising from $5s^2$ electrons of $Sb^{3+}$ ions may be visualized from the ELF plot. Figure 4e displays the plot of the ELF for the experimentally obtained centrosymmetric $Cs_3Sb_2Cl_9$ structure. We find that the electron density around the Sb is asymmetric and forms a usual lobe shape arising from the 5*s* lone-pair of Sb. We argued from the plot of the DOS that the hybridization with anion p orbitals (Cl-*p*) plays an important role in the formation of an asymmetric lobe shaped iso-surface of the ELF for the stereochemically active lone-pair promoting local distortion. These local distortions due to lone-pairs imparts the local optically anisotropic environment. It is however interesting to note that these lone-pairs do not create global structural distortion and the structure remains globally centrosymmetric, as the lone-pair lobes on pairs of Sb sites are arranged in an opposite manner.

The orientation of the $SbCl_6$ octahedra in the vacancy ordered $Cs_3Sb_2Cl_9$ crystal structure leads to the existence of multiple energy transition bands globally, which originate from the hybridization of the 5*s* orbital of Sb with 5*p* orbital of Sb and 3*p* orbital of Cl. The hybridization is further influenced by the bond angle of Sb−Cl−Sb along apical and equatorial directions globally. Larger bond angle deviations from plane geometry (i.e., bond angle of 180°) results in the weaker hybridization and lower transition states of CBM.[59] Our single crystal XRD measurements revealed that the bond angle of equatorial Sb−Cl−Sb in $Cs_3Sb_2Cl_9$ is larger by nearly 4° than that of apical one (Figure 1d). Hence, the lowest optical anisotropy is expected to occur involving mainly the orbitals hybridized along the equatorial direction compared to the apical direction. The change in the bond angle also produces distortion in the $SbCl_6$ octahedral arrangement in the crystal structure promoting the required anisotropy for the birefringence. It shows that exploring the contributions of microscopic units that form the perovskite framework is an efficient way to design novel materials with birefringence properties.

In conclusion, we report a new all-inorganic halide perovskite single crystal, $Cs_3Sb_2Cl_9$, which exhibits the birefringence of a comparable value as in commercial birefringent crystals. Vacancy ordered perovskite contains zigzag double chain $SbCl_6$ octahedral linkage in the orthorhombic phase. The classical Sb−Cl coordination octahedra is distorted owing to the stereochemical activity of lone-pair active Sb (III) cations. The highly distorted $SbCl_6$ polyhedra made the major contribution to the birefringence of $Cs_3Sb_2Cl_9$. Furthermore, the hybridization of Sb 5*s* with Sb 5*p* and Cl 3*p* states reinforces the stereochemical activity of lone-pair of the Sb (III) cations. Microscopically, the change in the Sb-Cl-Sb bond angle that introduces distortion in the $SbCl_6$ octahedral arrangement in the apical and equatorial directions within the crystal structure produces the required anisotropy for the birefringence. This work expands the frontiers of lone-pair cation-based halide perovskites. Taking into account the excellent optoelectronic properties of halide perovskites, the birefringent $Cs_3Sb_2Cl_9$ may also find significant applications in integrated devices.

ASSOCIATED CONTENT

**Supporting Information** Experimental details of the material synthesis at different condition, characterizations, birefringence measurement, single crystal XRD measurements and analyses of $Cs_3Sb_2Cl_9$ single crystal, powder XRD pattern and photographs of $Cs_3Sb_2Cl_9$ single crystals, details of TEM-EDS elemental analysis, TGA and DSC data, XPS survey spectrum, Tauc plots Corresponding to UV-Visible and Diffused reflectance spectroscopy, SEM image, computational details, stereochemical activity calculation, references.

AUTHOR INFORMATION

***Corresponding Author***

Prof. Somobrata Acharya−*School of Applied and Interdisciplinary Sciences and Technical Research Centre, Indian Association for the Cultivation of Science, Kolkata 700032, India*

*Orchid: 0000-0001-5100-5184; Email:* camsa2@iacs.res.in

AUTHOR ADDRESS

Ms. Shramana Guha − School of Applied and Interdisciplinary Sciences, Indian Association for the Cultivation of Science, Jadavpur, Kolkata-700032, India.

Dr. Amit Dalui − School of Applied and Interdisciplinary Sciences, Indian Association for the Cultivation of Science, Jadavpur, Kolkata-700032, India.

Dr. Piyush Kanti Sarkar – School of Applied and Interdisciplinary Sciences, Indian Association for the Cultivation of Science, Jadavpur, Kolkata-700032, India.

Dr. Sima Roy – School of Applied and Interdisciplinary Sciences, Indian Association for the Cultivation of Science, Jadavpur, Kolkata-700032, India.

Dr. Atanu Paul – School of Physical Sciences, Indian Association for the Cultivation of Science, Jadavpur, Kolkata-700032, India.

Prof. Indra Dasgupta – School of Physical Sciences, Indian Association for the Cultivation of Science, Jadavpure, Kolkata-700032, India.

Mr. Sujit Kamilya – Solid State and Structural Chemistry Unit, Indian Institute of Science, Sir C V Raman Road, Bangalore 560012, India.

Dr. Abhishake Mondal – Solid State and Structural Chemistry Unit, Indian Institute of Science, Sir C V Raman Road, Bangalore 560012, India. *http://m2ssscuiisc.in;*

Prof. D. D. Sarma – Solid State and Structural Chemistry Unit, Indian Institute of Science, Sir C V Raman Road, Bangalore 560012, India; Orchid: *0000-0001-6433-1069*

**Notes**

The authors declare no competing financial interest.

ACKNOWLEDGMENT

We acknowledge Science and Engineering Research Board (SERB) grant SERB-STAR grant STR/2020/000053 for financial support. We acknowledge Supriya Chakraborty, IACS for the support with the TEM measurements and Md. Yusuf Sk., IACS for the SEM measurements. AM thanks to the Council of Scientific and Industrial Research (CSIR, Project No: 01(3031)/21/EMR-II) and SK thanks CSIR for the research fellowship. SA and ID acknowledge Technical Research Centre (TRC), Department of Science & Technology for support.

**Supporting Information**

# Stereochemically Active Lone-pair Leads to Strong Birefringence in the Vacancy Ordered $Cs_3Sb_2Cl_9$ Perovskite Single Crystals

*Shramana Guha,[a] Amit Dalui,[a] Piyush Kanti Sarkar,[a] Sima Roy,[a] Atanu Paul,[b] Sujit Kamilya,[c] Abhishake Mondal,[c] Indra Dasgupta,[b] D. D. Sarma[c] and Somobrata Acharya[a,d]**

[a] School of Applied and Interdisciplinary Sciences, Indian Association for the Cultivation of Science, Jadavpur, Kolkata-700032, India.

[b] School of Physical Sciences, Indian Association for the Cultivation of Science, Jadavpur, Kolkata-700032, India.

[c] Solid State and Structural Chemistry Unit, Indian Institute of Science, Sir C V Raman Road, Bangalore 560012, India.

[d]Technical Research Center, Indian Association for the Cultivation of Science, Jadavpur, Kolkata-700032, India.

**Corresponding Author**

*E-mail: camsa2@iacs.res.in

**Table of Content:**

**$Cs_3Sb_2Cl_9$ single crystals synthesis:** Antimony chloride ($SbCl_3$, 98%, Sigma), cesium chloride (CsCl, 98%, Merck), hydrochloric acid (HCl, 37%, Merck) were used as received without further purifications. All experiments were performed under ambient condition. First, white precipitate of polycrystalline $Cs_3Sb_2Cl_9$ was synthesized by mixing 1 mL of 37% HCl solution of $SbCl_3$ (0.457 g, 2.0 mmol) with a 1 mL solution of CsCl (0.504 g, 3.0 mmol) in 37% HCl under stirring. After 60 min of vigorous stirring at room temperature, the single crystal of $Cs_3Sb_2Cl_9$ was obtained by re-crystallization of the above-mentioned polycrystalline precipitate in HCl. Large size single crystals were achieved by heating a dilute solution (0.05 g/ mL) of $Cs_3Sb_2Cl_9$ in concentrated HCl at 120°C in a glass vessel to make clear solution. Then, slow cooling (1 °C/ 10 minutes) of this solution over 10 hours leads to the formation of colourless rod-shaped $Cs_3Sb_2Cl_9$ crystals.

**Synthesis of hexagonal shaped single crystals:** First, white precipitate of polycrystalline $Cs_3Sb_2Cl_9$ was synthesized by mixing 1 mL of 37% HCl solution of $SbCl_3$ (0.457 g, 2.0 mmol) with a 1 mL solution of CsCl (0.504 g, 3.0 mmol) in HCl under stirring. After nearly 60 minutes of vigorous stirring at room temperature, the single crystal of $Cs_3Sb_2Cl_9$ was obtained by re-crystallization of the above-mentioned polycrystalline precipitate in HCl. Large size single crystals of hexagonal shape achieved by heating a dilute solution (0.04 g/ mL) of $Cs_3Sb_2Cl_9$ in concentrated HCl to 120°C in a glass vessel to make clear solution. Then, slow cooling (1 °C/ 10 minutes) of this solution over the course of 10 hours lead to the formation of colourless hexagonal crystals.

**Synthesis of chunk shaped single crystals:** First, white precipitate of polycrystalline $Cs_3Sb_2Cl_9$ was synthesized by mixing 1 mL of 37% HCl solution of $SbCl_3$ (0.457 g, 2.0 mmol) with a 1 mL solution of CsCl (0.504 g, 3.0 mmol) in HCl under stirring. After nearly 60 minutes of vigorous stirring at room temperature, the single crystal of $Cs_3Sb_2Cl_9$ was obtained by re-crystallization of the above-mentioned polycrystalline precipitate in HCl. The solution was diluted in (nearly 0.07 g/mL) concentrated HCl and heated at 120°C for some time. Then, slow cooling (1 °C/ 10 minutes) of this solution over the course of 10 hours lead to the formation of colourless crystalline chunks.

**Synthesis of irregular plate shaped single crystals:** First, white precipitate of polycrystalline $Cs_3Sb_2Cl_9$ was synthesized by mixing 1 mL of 37% HCl solution of $SbCl_3$ (0.457 g, 2.0 mmol) with a 1 mL solution of CsCl (0.504 g, 3.0 mmol) in HCl under stirring. After nearly 60 minutes of vigorous stirring at room temperature, the single crystal of $Cs_3Sb_2Cl_9$ was obtained by re-crystallization of the above-mentioned polycrystalline precipitate in HCl. Large size single crystals were achieved by heating a dilute solution (0.05 g/ mL) of $Cs_3Sb_2Cl_9$ in concentrated HCl to 120°C in a glass vessel to make clear solution. After 60 minutes, the temperature was turned off immediately and the solution was transferred to a chilled environment and kept for overnight crystallization.

**Characterizations:** Single crystal X-ray diffraction (XRD) data for the representative compound were collected using a Bruker SMART APEX II ULTRA diffractometer, equipped with a CCD type area detector with graphite-monochromated Mo-Kα radiation ($\lambda$ = 0.71073 Å). Powder XRD measurements were carried out using Rigaku SmartLab diffractometer using Cu Ka ($\lambda$ = 1.54 Å) as the incident radiation. Transmission electron microscopy (TEM) was performed using UHR-FEG-TEM, JEOL JEM-2010 electron microscopy operating at 200 kV electron source. Selected area electron diffraction (SAED) and energy dispersive x-ray spectroscopy (EDS) were also obtained with the same electron microscope. Finely grinded single crystals were dispersed in toluene and that dilute dispersion solution was placed on a carbon-coated copper grid and vacuum dried prior to measurements. Scanning electron microscopy (SEM) measurements were performed using a Zeiss Gemini 500 Field Emission Scanning Electron Microscope. Raman spectra were recorded using a J-Y HORIBA T64000 triple Raman spectrophotometer. Absorption spectra were obtained using Varian Carry 5000 UV−vis-NIR spectrophotometer. X-ray photoelectron spectroscopy (XPS) was measured g Perkin Elmer Phi 5500 ESCA spectrometer with an Al Kα X-ray source generating X-ray photons of 1,486.7 eV in energy in an ultrahigh-vacuum chamber with a base pressure of $10^{-9}$ Torr. The spectra were obtained with an analyzer pass-energy of 23.5 eV and a scan speed of 0.05 eV $s^{-1.}$ For XPS measurements, a $Cs_3Sb_2Cl_9$ single crystals was pasted on carbon tape on top of a silicon wafer and that whole substrate was placed in ultra-high vacuum of the instrument for measurement. Thermogravimetric analysis (TGA) experiment was performed with a Q600, TA Instruments.

**Birefringence measurements:** The single crystals were chemically exfoliated into the smaller sizes ranging from 3 to 20 μm. The small crystal was chosen to measure, in order to improve the accuracy of the birefringence measurements. All birefringence measurements were carried out using Olympus BX53M microscope containing compensator. The crystal was placed at the centre of the field of view. First, the crystal was moved into the extinction position by rotating the object stage. Then, the object stage was rotated by ± 45° so that the crystal appeared bright from the dark (diagonal position). At this position, the compensator was introduced into the slot of the microscope. The compensator was rotated both in the clockwise and anti-clockwise by rotating the compensator drum until the crystal appeared dark (extinction position). The tilting angle was recorded as I′. The measurement was repeated by rotating the compensator in the opposite direction and the tilting angle was recorded as I″ at the extinction position. The average tilting angles were calculated using $I = (I' - I'')/2$. Simultaneously, Michel-Lévy interference colour chart was monitored. The phase difference i.e., retardation (R) was obtained from the reference compensator variable U-CBE chart specification using the measured average tilting angle. The value of retardation from the table of compensator was found to be 830.14 μm. Thickness of the chemically exfoliated single crystals was determined from SEM (figure S8). The average thickness of the chemically exfoliated single crystals was found to be 6.91 μm. The birefringence was calculated using following formula:

$$R = | n_e - n_o | \times d = \Delta n \times d$$

where Δn represents the difference between refractive indices of extraordinary and ordinary ray; i.e., the value of birefringence, R is the retardation and d is the average thickness of the exfoliated crystallites. The value of birefringence is calculated to be $\Delta n = 0.12 \pm 0.01$ at 550 nm.

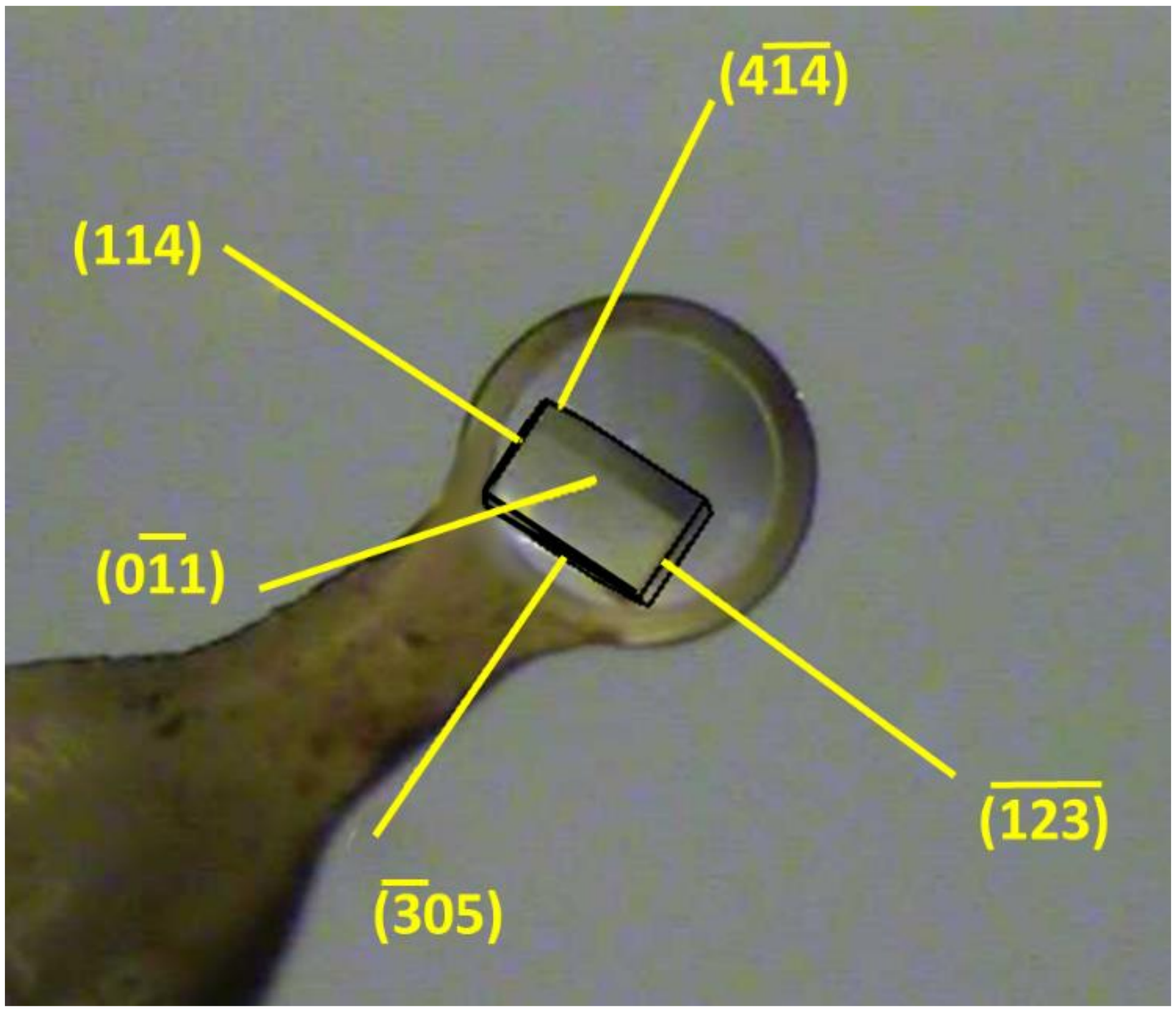

**Figure S1.** SCXRD measurement on top of the planar surface of a single crystal revealing (114) lattice planes perpendicular the elongated surface of the rod.

**Table S1.** Single crystal X-ray diffraction crystallographic table of $Cs_3Sb_2Cl_9$ crystals

| Compound | $Cs_3Sb_2Cl_9$ |
|---|---|
| CCDC Number | 2288964 |
| Empirical formula | $Cs_6Sb_4Cl_{18}$ |
| Formula weight /g mol$^{-1}$ | 1922.56 |
| Temperature / K | 293 (2) |
| Crystal system | Orthorhombic |
| Space group | *P*nma |
| a, Å | 18.6521 (8) |
| b, Å | 7.5774 (4) |
| c, Å | 13.0625(5) |
| $\alpha$, ° | 90 |
| $\beta$, ° | 90 |
| $\gamma$, ° | 90 |
| V, Å$^3$ | 1846.18 (14) |
| Z | 2 |
| $d_{cal}$, g cm$^{-3}$ | 3.458 |
| $\mu$, mm$^{-1}$ | 10.028 |
| F(000) | 1680 |
| 2Θ range for data collection/° | 6.928 to 59.188 |
| Index ranges | $-22 \leq h \leq 25$, $-8 \leq k \leq 10$, $-17 \leq l \leq 16$ |
| Completeness | 0.91 |
| Reflections collected | 13458/2518 |
| Independent reflections | 2518 [$R_{int}$ = 0.0581, $R_{sigma}$ = 0.0415] |
| Data/restraints/parameters | 2518/0/77 |
| Goodness-of-fit on $F^2$ | 1.066 |
| Final R1 indices [I>2σ(I)] | $R_1$ = 0.0379, $wR_2$ = 0.0814 |
| Final wR2 indices [all data] | $R_1$ = 0.0532, $wR_2$ = 0.0920 |
| Largest diff. peak/hole / e Å$^{-3}$ | 1.28/-1.51 |

$R1 = \sum||Fo| - |Fc||/\sum|Fo|$ and $wR2= |\sum w(|Fo|^2 - |Fc|^2 )|/\sum|w(Fo)^2|^{1/2}$

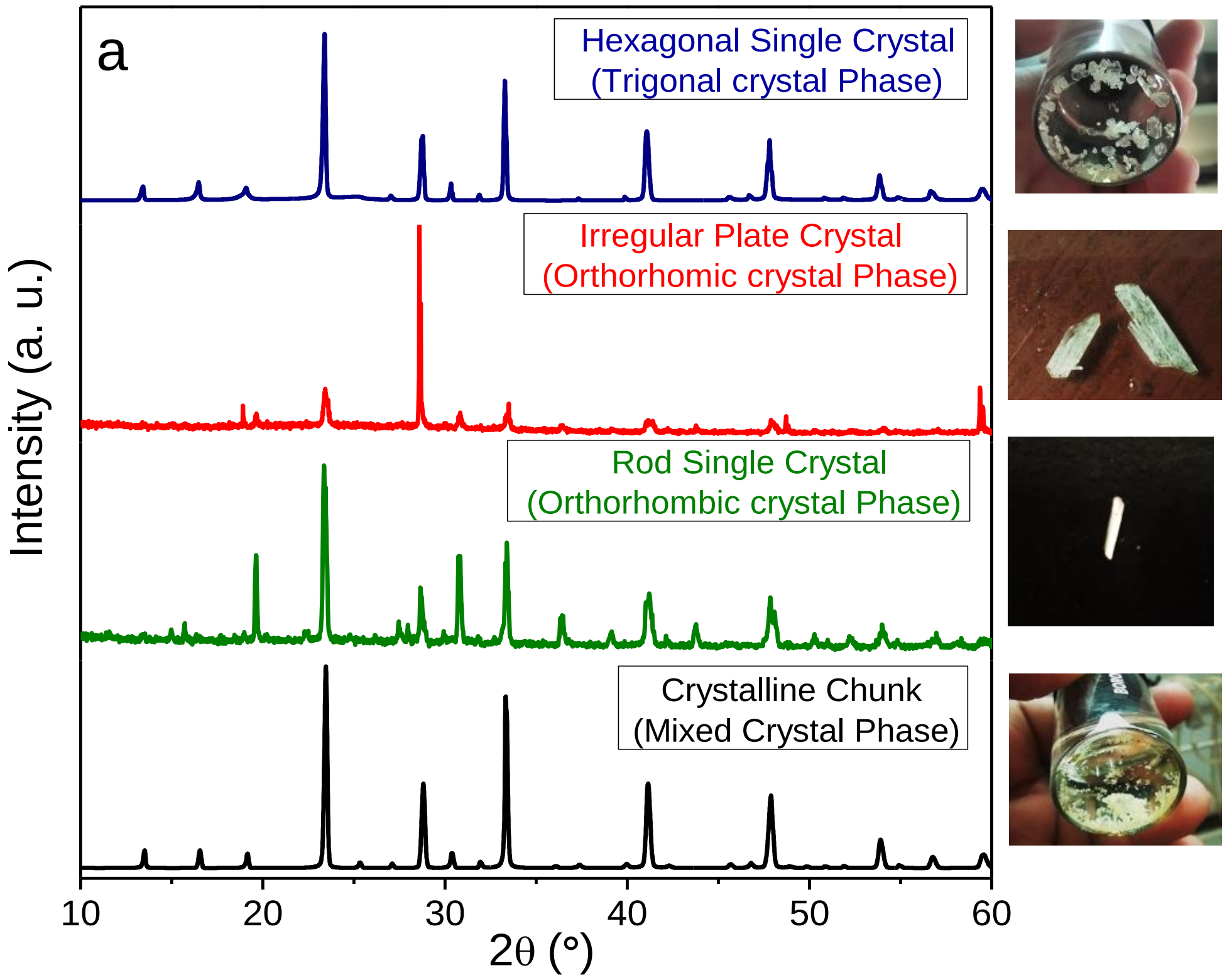

**Figure S2.** Powder XRD pattern and photographs of $Cs_3Sb_2Cl_9$ single crystals grown from different concentrations of the precursor solution of cesium and antimony chloride in concentrated HCl, as well as under different heating condition in some cases.

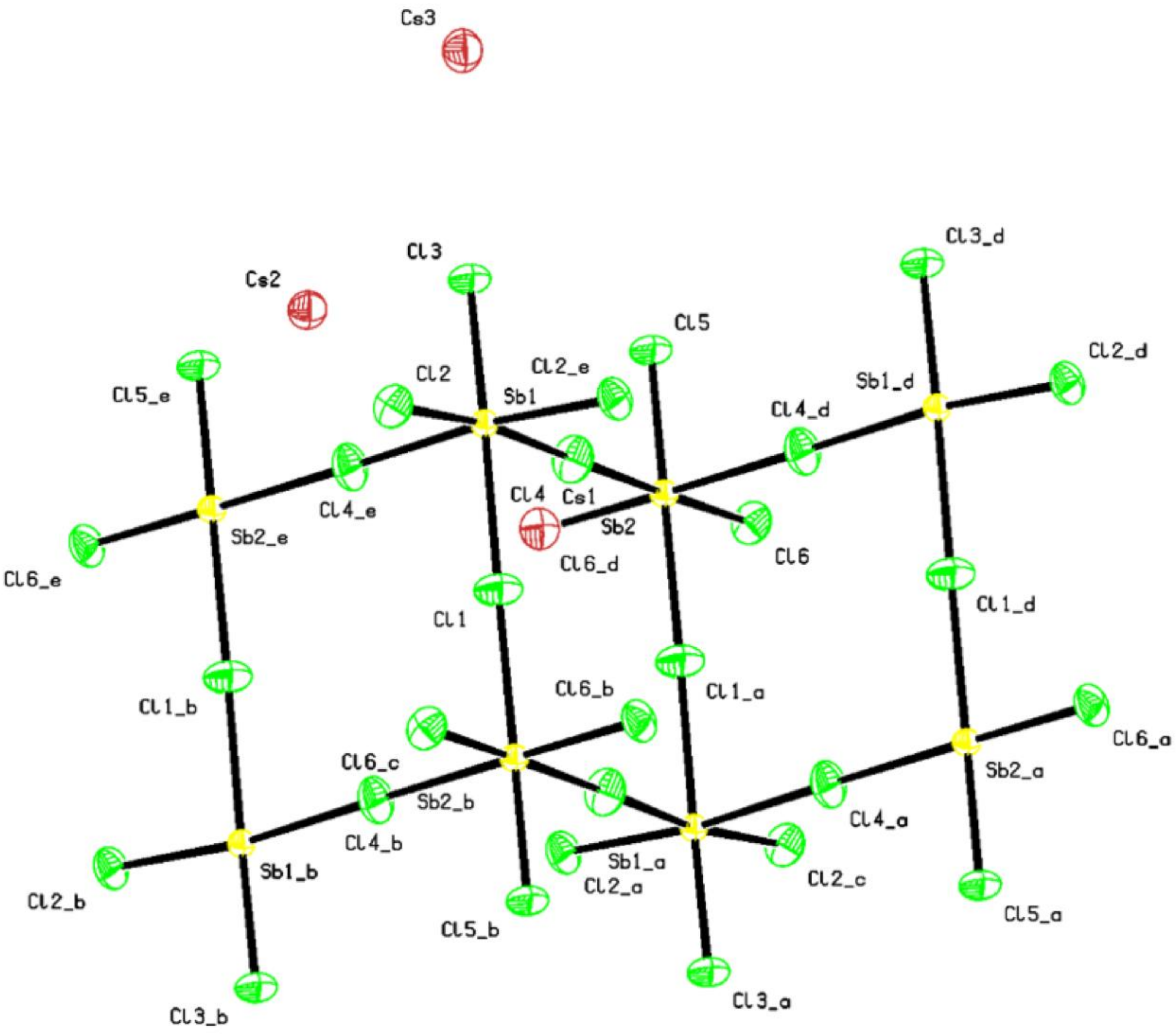

**Figure S3.** Structure analyses of $Cs_3Sb_2Cl_9$ single crystals.

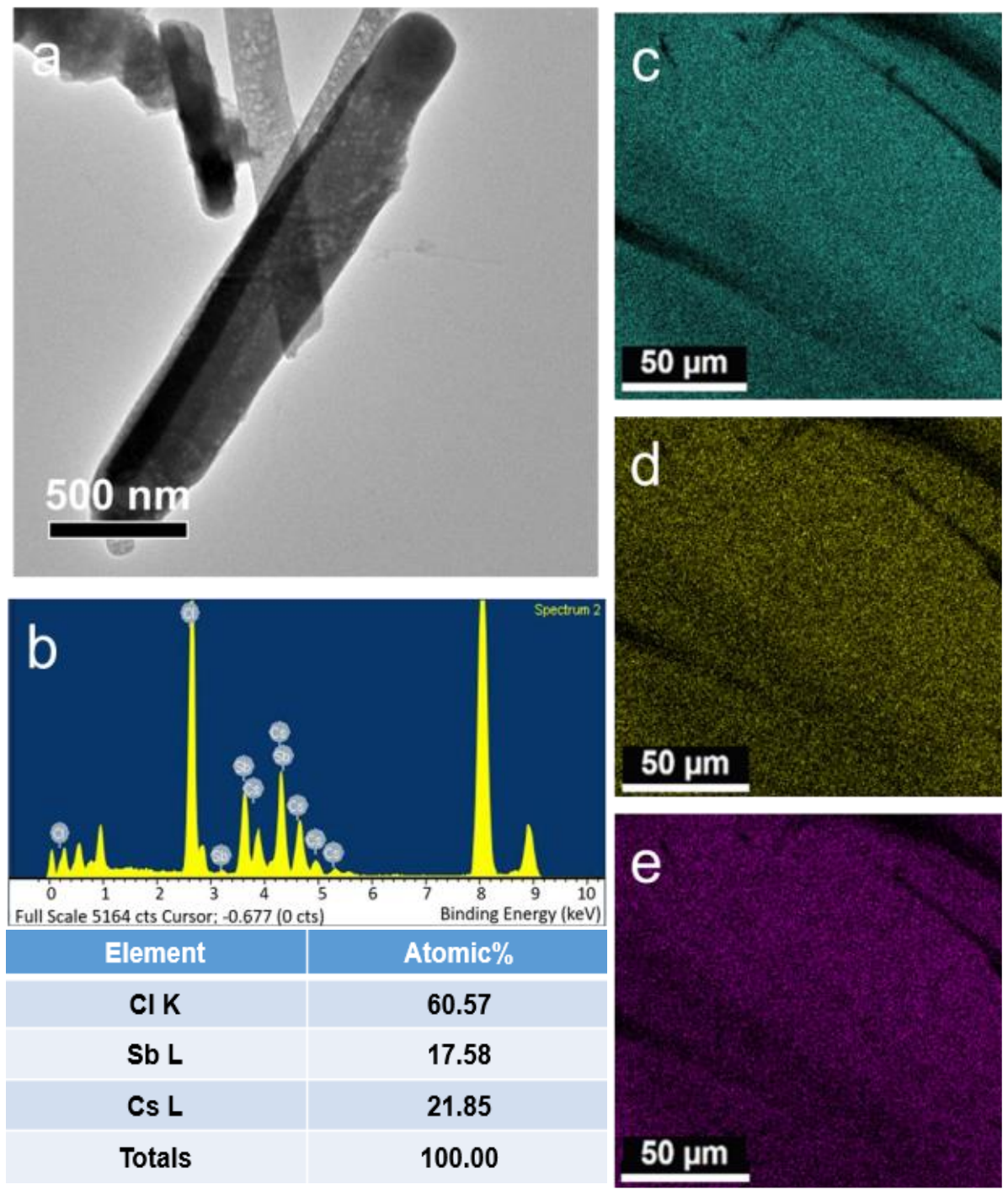

| Element | Atomic% |
|---|---|
| Cl K | 60.57 |
| Sb L | 17.58 |
| Cs L | 21.85 |
| Totals | 100.00 |

**Figure S4.** TEM analyses of finely grinded $Cs_3Sb_2Cl_9$ single crystals. (a) TEM image. (b) EDS spectrum revealing the elemental analyses EDS spectrum shows a ratio of 3:2:7 of Cs:Sb:Cl**.** (c-e) HAADF-STEM image and EDS elemental mapping showing homogeneous distribution of Cs, Sb, and Cl elements.

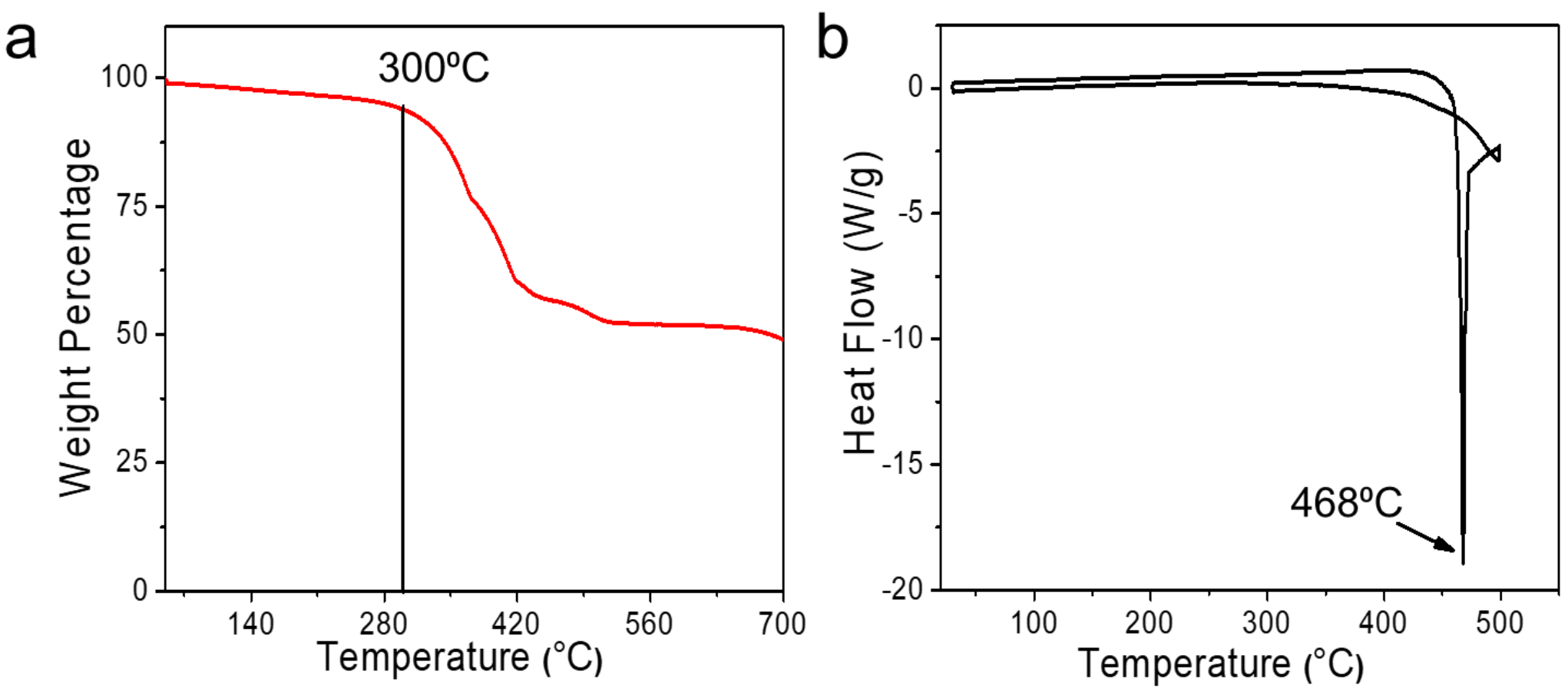

**Figure S5.** (a) Thermogravimetric analysis curve of $Cs_3Sb_2C1_9$ single crystals. (b) Differential scanning calorimetry curve of $Cs_3Sb_2C1_9$ single crystals.

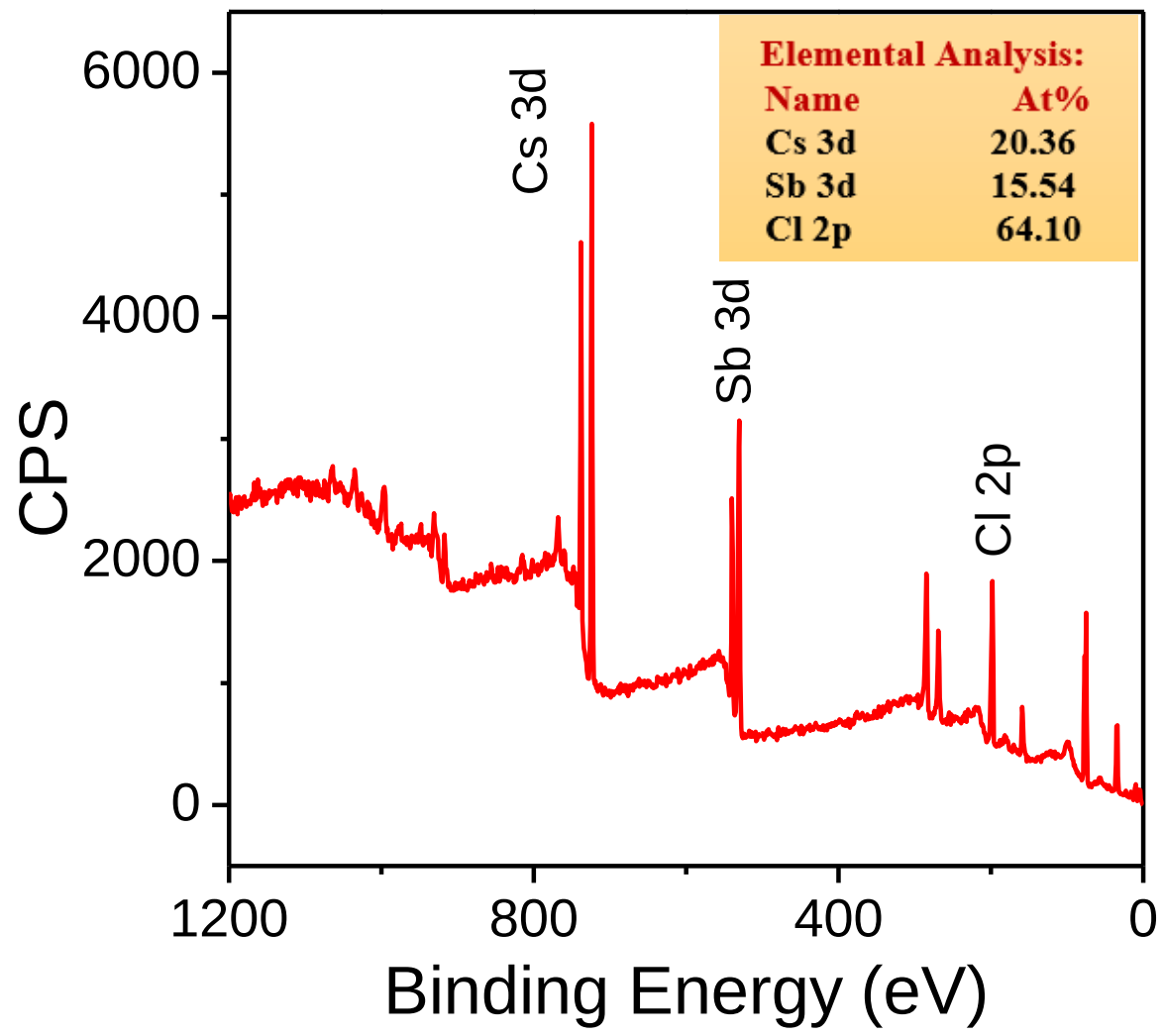

**Figure S6.** X-ray photoelectron spectroscopy survey spectrum of $Cs_3Sb_2Cl_9$ single crystals.

XPS survey spectra of the crystal was quantified in terms of peak intensities and peak positions. The peak intensities help to measure quantity of an element at the surface, while the peak positions indicate the elemental and chemical composition. We have used CasaXPS software for performing the elemental analysis of the single crystal, where the Cs 3*d* orbital, Sb 3*d* orbital and Cl 2*p* orbital was considered during the procedure. Area under the curve of Cs 3*d*, Sb 3*d* and Cl 2*p* XPS spectra were divided by corresponding XPS cross-section value to obtain final ratio which indicates the elemental composition of the single crystal. XPS spectra shows a relative ratio of Cs:Sb:Cl ≈ 2.5:2:8 which nearly matches with the EDS measurements.

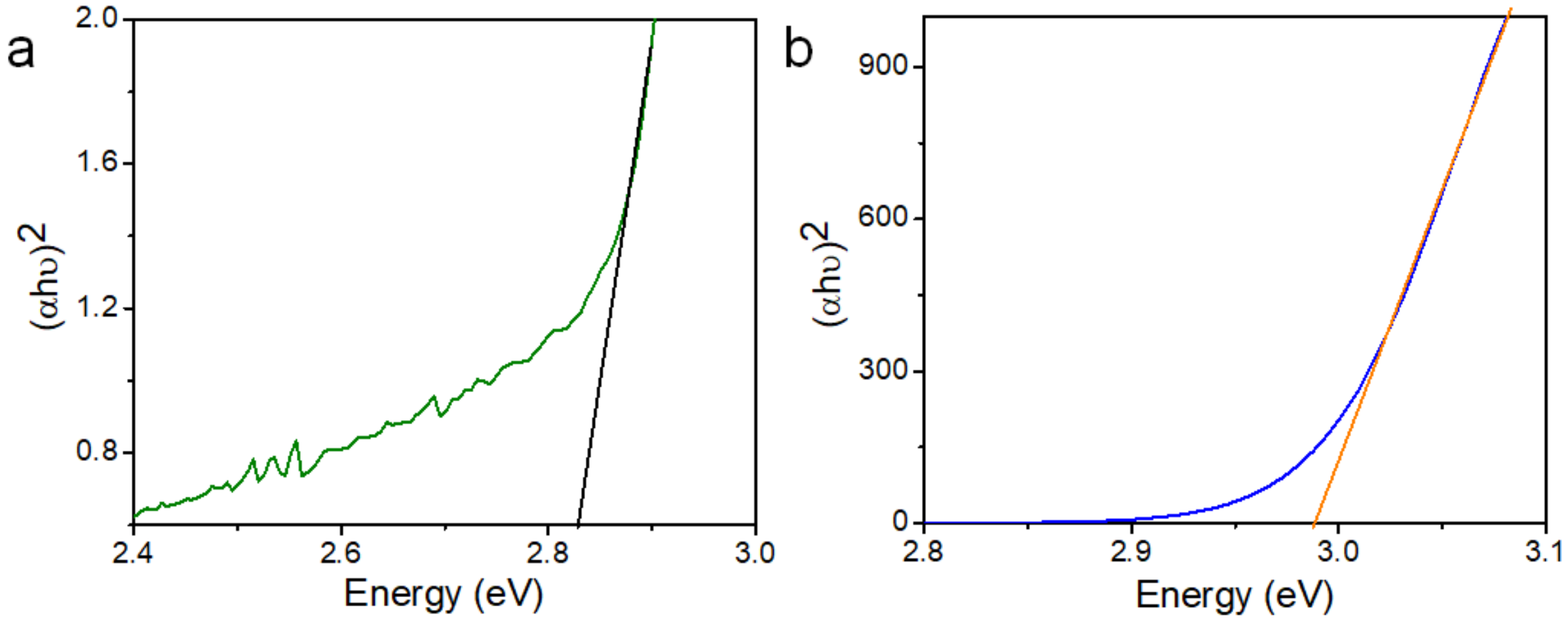

**Figure S7.** Tauc plot from (a) UV-Visible Spectroscopy showing the band gap of 2.86 eV (b) Diffuse Reflectance Spectroscopy, showing the band gap of 2.98 eV of the $Cs_3Sb_2Cl_9$ single crystals.

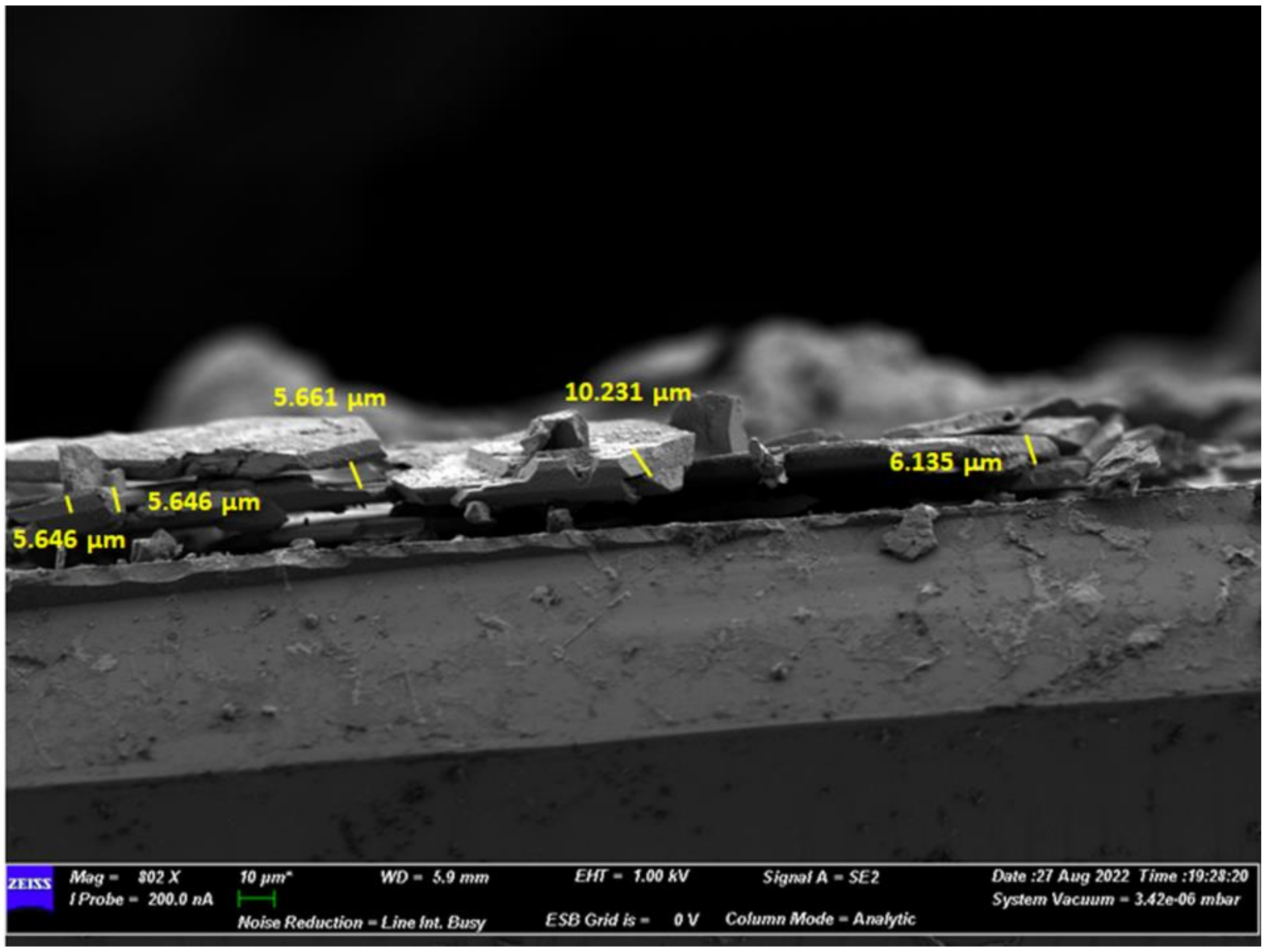

**Figure S8.** SEM image of the $Cs_3Sb_2Cl_9$ crystallites showing an average thickness of 6.91 µm.

**Computational Details:** The first-principles electronic structure calculations were performed in the framework of density functional theory (DFT)[S1,S2] using the plane-wave basis set supplemented with projector augmented wave (PAW)[S3] method as implemented in the Vienna ab initio simulation package (VASP)[S4,S5] where the generalized gradient approximation (GGA) by Perdew-Burke-Ernzerhof (PBE)[S6] formalism was used for the exchange correlation functional. The Brillouin-zone integration was carried out by considering 4×8×4 k-point mesh. The plane wave cut-off energy was set to 600 eV. The valence electron configuration for the PAW potentials were Cs: [core] $5s^2$ $5p^6$ $6s^1$, Sb: [core] $5s^2$ $5p^3$, Cl: [core] $3s^2$ $3p^5$. The maximally-localized Wannier functions were calculated by WANNIER90 code[S7] as interfaced with VASP.

The frequency-dependent optical properties were calculated using the independent particle approximation as implemented in VASP.

**Calculation of stereochemical activity:** The degree of stereochemical activity of lone pair is measured by calculating the difference between the Sb−Cl bond distances from hemispheres I and II (Figure 4b). The bond distances were taken from the single crystal X-ray measurement data. Here, $\Delta d_1$ = 0.003 Å and $\Delta d_2$ = 0.012 Å, which specify a certain degree of stereochemical activity of the Sb (III) lone pair in the crystal structure.